\documentclass[a4paper,12pt]{article}

\newcommand{\sect}[1]{\setcounter{equation}{0}\section{#1}}

\begin{document}

\topmargin 0pt
\oddsidemargin 0mm

\renewcommand{\thefootnote}{\fnsymbol{footnote}}

\begin{titlepage}
\begin{flushright}
OU-HET 355 \\
hep-th/0007106
\end{flushright}

\vspace{8mm}
\begin{center}
{\Large\bf (F1, D1, D3) Bound State, Its Scaling Limits and
$SL(2, {\bf Z})$ Duality}
\vspace{15mm}

{\large
Rong-Gen Cai\footnote{e-mail address: cai@het.phys.sci.osaka-u.ac.jp} and
Nobuyoshi Ohta\footnote{e-mail address: ohta@phys.sci.osaka-u.ac.jp}} \\
\vspace{10mm}
{\em Department of Physics, Osaka University,
Toyonaka, Osaka 560-0043, Japan}

\end{center}
\vspace{15mm}
\centerline{{\bf{Abstract}}}
\vspace{5mm}

We discuss the properties of the bound state (F1, D1, D3) in IIB
supergravity in three different scaling limits and the $SL(2,{\bf Z})$
transformation of the resulting theories. In the simple decoupling limit
with finite electric and magnetic components of NS $B$ field, the worldvolume
theory is the ${\cal N}$=4 super Yang-Mills (SYM) and the supergravity dual
is still the $AdS_5 \times S^5$. In the large magnetic field limit with
finite electric field, the theory is the noncommutative super Yang-Mills
(NCSYM), and the supergravity dual is the same as that without the electric
background. We show how to take the decoupling limit of the closed string
for the critical electric background and finite magnetic field, and that
the resulting theory is the noncommutative open string (NCOS) with both
space-time and space-space noncommutativities. It is shown that under the
$SL(2, {\bf Z})$ transformation, the SYM becomes itself with a different
coupling constant, the NCSYM is mapped to a NCOS, and the NCOS in general
transforms into another NCOS and reduces to a NCSYM in a special case.

\end{titlepage}

\newpage
\renewcommand{\thefootnote}{\arabic{footnote}}
\setcounter{footnote}{0}
\setcounter{page}{2}


\sect{Introduction}

Over the past few years, one of the important progresses in string and field
theories is the observation of Maldacena~\cite{Mald1} that open string
excitations on D-branes decouple from gravity in the appropriate low energy
limit, the so-called decoupling limit and that string/M theories on the
anti-de Sitter space (AdS) are dual to a certain large $N$ conformal field
theory (CFT) which lives on the boundary of the AdS. One important example
of this AdS/CFT correspondence is that IIB string theory on the
$AdS_5 \times S^5$ is believed to be dual to the ${\cal N}$=4 super Yang-Mills
theory (SYM). The geometry $AdS_5 \times S^5$ comes from the decoupling
limit of D3-branes (hereafter referred to as the SYM limit). Through this
``duality'' one can learn something about the large $N$ SYM with strong
't Hooft coupling from the low energy limit of the superstring/M theory,
supergravity. Indeed a lot of knowledge has been acquired from the
correspondence, which is consistent with our expectation.

Recently it has been noticed that when a constant NS $B$ field is present,
the worldvolume coordinates of D-branes become
noncommutative~\cite{Chu,Ard,Seiberg}. In an appropriate limit the worldvolume
theory of D-branes with nonvanishing spatial components of the NS $B$ field
is a noncommutative SYM (NCSYM)~\cite{Mald2}-\cite{LS}
(hereafter called the NCSYM limit) with space-space noncommutativy.
More recently when an electric background is introduced, the resulting theory
has been found to be a noncommutative open string (NCOS) theory with
space-time noncommutativity in an appropriate limit
(hereafter NCOS limit)~\cite{Seiberg2,Gop1}. For related discussions, see
also Refs.~\cite{Seib3}-\cite{AGM}.

In this paper we consider the above three limits for the bound state
(F1, D1, D3) with both electric and magnetic $B$ fields~\cite{Lu1,Lu2,Lu3}.
The SYM limit for this bound state has been discussed in Ref.~\cite{Lu3}. For
completeness and to compare with the other two cases, we also review the SYM
limit in this paper. In the simple $\alpha'\to 0$ limit, the resulting
theory is still the ${\cal N}$=4 SYM in ordinary spacetime. In the NCSYM
limit the theory is the NCSYM with space-space noncommutativity again,
namely without space-time noncommutativity. In the NCOS limit the
resulting theory is most nontrivial and is the NCOS with not only space-time
noncommutativity but also space-space noncommutativity. Another subject we
discuss is the $SL(2, {\bf Z})$ duality of these solutions. The S-duality
has been discussed for theories with either an electric or magnetic
background~\cite{Gop1}, but full understanding of the behavior of these
theories with both types of backgrounds under the general $SL(2,{\bf Z})$
duality has not been obtained. In the NCOS theory with both space-time and
space-space noncommutativities, there remains in general a nonvanishing axion
field and it seems  more appropriate to use the $SL(2, {\bf Z})$ transformation
to get clear understanding of the relations of these theories. We
show that these theories are nicely related with each other by the
$SL(2, {\bf Z})$ transformation of IIB supergravity.

As a preparation for our following discussions, in sect.~2 we review the
Seiberg-Witten relation between the open and closed string moduli when both the
electric and magnetic components of the NS $B$ field are present. In sect.~3
we discuss supergravity duals for three different limits. In particular we
show how we are uniquely lead to the nontrivial decoupling limit for
space-time and space-space noncommutativities. In sect.~4, we discuss general
$SL(2, {\bf Z})$ transformations of these solutions and show that under
the $SL(2,{\bf Z})$ transformation, the NCSYM is mapped into a
NCOS with space-time and space-space noncommutativities, but the NCOS
transforms in general into another NCOS with different
parameters and reduces to a NCSYM in a special case, depending on the
asymptotic value of the axion. Concluding remarks are given in sect.~5.

In the course of our work, two papers~\cite{LRS,RS} have appeared in which
similar topics were discussed.
The authors of Ref.~\cite{LRS} have discussed, in the setting of a vanishing
asymptotic value of the axion, the S-duality of the NCOS theory and NCSYM,
from both points of view of string theory and supergravity.
In Ref.~\cite{RS}, the authors have discussed the $SL(2,{\bf Z})$
transformation of NCOS from the open
string point of view.\footnote{More recently, related discussion on the
S-duality of NCSYM  has been given in Refs.~\cite{KT,RU}.}
Our discussion is in terms of the dual supergravity description.

\sect{ Seiberg-Witten relation}

When a constant NS $B$ field is present, an open string ending on a
D-brane has the following boundary conditions:
\begin{equation}
\label{boundary}
g_{ij}\partial_{\sigma}X^j +2\pi \alpha'B_{ij}\partial_{\tau}X^{j}=0,\ \ \
  \delta X^a=0.
\end{equation}
The open string moduli appear in the disk correlators on the open string
worldsheet boundaries
\begin{equation}
\langle X^i(\tau)X^j(0)\rangle = -\alpha' G^{ij} \ln(\tau)^2
 +\frac{i}{2}\Theta ^{ij} \epsilon (\tau).
\end{equation}
The open and closed string moduli are connected by the Seiberg-Witten
relation~\cite{Seiberg}
\begin{eqnarray}
&& G_{ij}=g_{ij}-(2\pi\alpha')^2 (Bg^{-1}B)_{ij}, \nonumber\\
&& \Theta^{ij}=2\pi \alpha' \left(\frac{1}{g+2\pi\alpha' B}
   \right)^{ij}_A, \nonumber\\
&& G^{ij}=\left(\frac{1}{g + 2\pi \alpha' B}\right)_S^{ij},\nonumber\\
&& G_s =g_s\left (\frac{\det G_{ij}}
   {\det(g_{ij} +2\pi\alpha' B_{ij})}\right)^{1/2},
\end{eqnarray}
where $(\;)_A$ and $(\;)_S$ denote the antisymmetric and symmetric parts,
respectively.

The constant NS $B$ field is equivalent to a gauge field on the worldvolume
of the D-brane because only ${\cal F}_{ij}=  B_{ij}+F_{ij}$ is gauge
invariant. Therefore the electric and magnetic components of the $B$ field
can always be rotated so that they are parallel to each other. On the other
hand, we are mainly concerned with the D3-brane case in this paper,
restricting ourselves to the worldvolume of D3-branes. Suppose we have the
closed string metric
\begin{equation}
g_{ij}=g_1(\delta_i^1\delta_j^1-\delta_i^0\delta_j^0) +g_2(\delta_i^2\delta_j^2
 +\delta_i^3\delta_j^3),
\end{equation}
and the constant $B$ field has components
\begin{equation}
B_{ij}=E(-\delta_i^1\delta_j^0 + \delta_i^0\delta_j^1)
   + B(\delta_i^2\delta_j^3-\delta_i^3\delta_j^2).
\end{equation}
Defining
\begin{equation}
e=\frac{E}{E_{\rm crit}}, \ \ \ b=\frac{B}{B_0},
\end{equation}
where
\begin{equation}
E_{\rm crit}=\frac{g_1}{2\pi\alpha'},\ \ \ B_0=\frac{g_2}{2\pi\alpha'},
\end{equation}
and using the Seiberg-Witten relation, we have the open string metric
\begin{equation}
\label{om}
G^{ij}= \frac{1}{g_1(1-e^2)}(-\delta^i_0\delta^j_0 +\delta_1^i\delta_1^j)
   +\frac{1}{g_2(1+b^2)}(\delta^i_2\delta^j_2 +\delta^i_3\delta^j_3),
\end{equation}
the noncommutativity matrix
\begin{equation}
\label{theta}
\Theta ^{ij}=\frac{2\pi\alpha' e}{g_1(1-e^2)}(\delta_0^i\delta_1^j-\delta^i_1
  \delta_0^j)
+\frac{2\pi\alpha'b}{g_2(1+b^2)}(-\delta_2^i\delta^j_3
+\delta^i_3\delta^j_2),
\end{equation}
and the open string coupling constant
\begin{equation}
\label{os}
G_s=g_s\sqrt{(1-e^2)(1+b^2)}.
\end{equation}

Let us now consider three different limits.

(1) {\it SYM limit.} From Eqs.~(\ref{om}), (\ref{theta}) and (\ref{os}), we see
that when $\alpha' \to 0$ with finite electric $E$ and magnetic components
$B$, the open and closed string moduli are equal. The noncommutativity matrix
$\Theta^{ij}$ vanishes, so that the entire spacetime becomes an ordinary
commutative one. In this case, the oscillation modes of an open string and
gravity are decoupled, and the worldvolume theory is the ${\cal N}$=4 SYM
in the low energy limit. Note that the constant $B$ field is converted into a
constant part of the gauge field on the worldvolume of the D3-brane and
remains in that limit. If the D3-brane worldvolume is noncompact, the constant
part of the gauge field is physically unmeasurable in the flat infinite
space~\cite{Gop1}. On the other hand, if the worldvolume is compact, the
constant part is quantized and the resulting low energy theory is the
${\cal N}$=4 SYM with both quantized electric and magnetic fluxes.

(2) {\it NCSYM limit.} Taking the limit $\alpha' \to 0$ with $g_1=1$, $g_2=
(2\pi \alpha' B)^2$, and $g_s= 2\pi\alpha' B G_s$ while keeping $G_s$,
$E$ and $B$ finite, we can obtain
\begin{equation}
G^{ij}=\eta^{ij}, \ \ \  \Theta ^{ij} =\frac{1}{B}(-\delta^i_2\delta^j_3
 +\delta^i_3\delta^j_2).
\end{equation}
In this case, the resulting theory is the noncommutative SYM (because
$\alpha' G^{ij}=0$, that is, massive open string modes are decoupled) with
space-space noncommutativity $\Theta^{23}\ne 0$; the Yang-Mills coupling
constant is $g_{YM}^2= 2\pi G_s$. The magnetic background gives rise to the
space-space noncommutativity. The constant electric component of the NS $B$
field is converted into a constant electric part of the gauge field, which is
physically unmeasurable if the worldvolume is noncompact again. If the
worldvolume is compact, the resulting theory is the NCSYM with quantized
electric flux.

(3) {\it NCOS limit.} In contrast to the magnetic component $B$, on which
there is no restriction, the electric component
cannot be beyond its critical value $E_{\rm crit}$. When the electric field
approaches its critical value in a certain manner, one may obtain a
noncritical NCOS theory, from which the closed string sector is
decoupled~\cite{Seiberg2,Gop1}, in spacetime with the space-time
noncommutativity. If a finite magnetic component is also present, the
space-space coordinates are also noncommutative. This has been noted in
Refs.~\cite{Chen} and \cite{LRS}. For example, taking the scaling
limit\footnote{Here $n$ is a positive parameter. It was chosen to be 2 in the
S-duality consideration in Ref.~\cite{Gop1}, and 1 in Ref.~\cite{Gop2}. We
will show in the next section that this freedom is allowed in this limit.}
\begin{eqnarray}
&&e = 1-\alpha'^n e_0/2, \qquad (n>0), \nonumber\\
&&g_1 =\frac{1}{\alpha'_{\rm eff}e_0\alpha'^{n-1}},\qquad
g_2=\frac{2\pi \alpha'b}{(1+b^2)\theta_1}, \nonumber\\
&&g_s=\frac{G_s}{\alpha'^{n/2}\sqrt{e_0(1+b^2)}},
\label{ncos}
\end{eqnarray}
with $e_0$ constant, while keeping $G_s$ and $B$ constant, one may get
\begin{equation}
\frac{\alpha'}{\alpha'_{\rm eff}}G^{ij}= - (\delta^i_0\delta^j_0
 -\delta^i_1\delta^j_1) +\frac{\theta_1}{\theta_0 b}(\delta^i_2\delta^j_2
 +\delta^i_3\delta^j_3),
\end{equation}
\begin{equation}
\Theta ^{ij}=\theta_0(\delta^i_0\delta^j_1 -\delta^i_1\delta^j_0)
 +\theta_1(-\delta^i_2\delta^j_3+\delta^i_3\delta^j_2),
\end{equation}
where $\alpha'_{\rm eff}=\theta_0/(2\pi)$. In this case, the critical electric
field is given by $E_{\rm crit}=\frac{1}{e_0 \theta_0\alpha'^n}$. Although the
closed string coupling constant is divergent in this limit, the open string
metric and coupling constant are well defined.

It is rather nontrivial to derive the dual gravity description of this
NCOS theory. One of our purposes in this paper is to elucidate this
problem. This is discussed in the next section.


\sect{Supergravity duals}

When both the electric and magnetic components of the NS $B$ field are
present on a D3-brane, the D3-brane becomes a (F1, D1, D3) bound state.
The supergravity configuration of the bound state has been constructed
in Ref.~\cite{Lu3}. It can also be constructed as follows~\cite{Mald2}.
Starting from a D3-brane without an NS $B$ field with
worldvolume coordinates ($x_0$, $x_1$, $x_2$ and $x_3$), and making a
T-duality along $x_3$, one gets a D2-brane with a smeared coordinate $x_3$.
Uplifting the D2-brane yields an M2-brane in the  11-dimensional supergravity.
Performing a coordinate rotation with parameter angles $\varphi$ and
$\theta$
\begin{eqnarray}
&& x_4=x_4'\cos\varphi - (x_2'\cos\theta -x_3'\sin\theta)\sin\varphi,
 \nonumber\\
&& x_2= x_4'\sin\varphi +(x_2'\cos\theta -x_3'\sin\theta)\cos\varphi,
 \nonumber\\
&& x_3= x_2'\sin\theta +x_3'\cos\theta,
\end{eqnarray}
and then reducing along $x_4'$, one obtains a new D2-brane. Applying
T-duality along $x_3'$, one reaches a (F1, D1, D3) bound state.
For the case of black configurations, the procedure is applicable as well.

Using the above approach, we obtain the non-extremal supergravity solution
of (F1, D1, D3) bound state\footnote{In this section a factor
$2\pi \alpha'$ is absorbed into the $B$ field.}$^,$\footnote{There is the
freedom to shift the asymptotic value of the axion field $\chi_\infty$
by the $SL(2,{\bf R})$ symmetry of the IIB supergravity. We note that this
is a classical symmetry of the IIB supergravity.}
\begin{eqnarray}
\label{solution}
&&ds^2= F^{-1/2}[h'(-f dx_0^2 +dx_1^2) +h(dx_2^2 +dx_3^2)]
    +F^{1/2}[f^{-1}dr^2 +r^2 d\Omega^2_5], \nonumber\\
&& e^{2\phi}=g^2_s hh', \ \ \
 \chi= - \frac{1}{g_sF}\tan\varphi \sin\theta, \nonumber \\
&& B_{01}=H^{-1}\coth\alpha \sin\varphi, \ \ \
 A_{01}= - H^{-1}\coth\alpha \sin\theta/g_s \cos\varphi, \nonumber\\
&& B_{23}= \frac{\tan\theta}{G},  \ \ \
 A_{23} =\frac{\tan \varphi \cos\theta}{g_sG}, \nonumber\\
&& F_{0123r} = \frac{\coth\alpha \cos\theta}{g_s \cos\varphi} h h'
 \partial_r F^{-1},
\end{eqnarray}
where
\begin{eqnarray}
&& f= 1-\frac{r_0^4}{r^4}, \ \ \
 H= 1+\frac{r_0^4\sinh^2\alpha}{r^4}, \nonumber\\
&& h= F/G, \ \ \  h'= F/H, \nonumber\\
&& F =1+\cos^2\varphi\frac{r_0^4 \sinh^2\alpha}{r^4},\ \ \
 G = 1+\cos^2\varphi\cos^2\theta\frac{r_0^4\sinh^2\alpha}{r^4}.
\end{eqnarray}
The bound state solution includes several special cases: when $\varphi =\theta
=0$, it reduces to the D3-brane solution; when $\varphi =\pi/2 $ and $\theta$
is arbitrary, it goes to the F-string solution with two smeared coordinates;
when $\varphi =0$ and $\theta =\pi/2$, it becomes the D-string solution
with two smeared coordinates; when $\varphi=0$ and $\theta$ is arbitrary, the
solution reduces to the (D1, D3) bound state;  when $\theta=0$ and
$\varphi$ is arbitrary, it becomes the (F1, D3) bound state; and finally
when $\theta =\pi/2$ and $\varphi$ is arbitrary, it goes back to the (F1, D1)
bound state with two smeared coordinates.

Some thermodynamic quantities, the ADM mass $M$, the Hawking temperature $T$,
and the entropy $S$, associated with the solution (\ref{solution})
are
\begin{eqnarray}
&& M= \frac{5\pi^3r_0^4 V_3}{16\pi G_{10}}(1+\frac{4}{5}\sinh^2\alpha),
 \nonumber\\
&& T= \frac{1}{\pi r_0\cosh\alpha}, \nonumber\\
\label{thermo}
&& S =\frac{\pi^3 r_0^5V_3}{4G_{10}}\cosh\alpha,
\end{eqnarray}
where $V_3$ is the spatial volume of worldvolume of the bound state.
An important feature of these thermodynamic quantities is their independence
of the parameter angles $\varphi$ and $\theta$. This means that the
thermodynamics is the same for all special cases discussed above. It also
guarantees the thermodynamic equivalence among the three theories coming from
different scaling limits which will be discussed shortly.

For the (F1, D1, D3) bound state, the charges of the three kinds of branes are
\begin{eqnarray}
&& Q_{D3}=\frac{4\pi^3 V_3 r_0^4 \sinh\alpha\cosh\alpha}{16\pi G_{10}}
   \cos\varphi \cos\theta, \nonumber \\
&& Q_{D1}=\frac{4\pi^3 V_3 r_0^4\sinh\alpha\cosh\alpha}{16\pi G_{10}}
    \cos\varphi \sin\theta, \nonumber\\
&& Q_{F}=\frac{4\pi^3 V_3 r_0^4\sinh\alpha\cosh\alpha}{16\pi G_{10}}
     \sin\varphi.
\end{eqnarray}
In the extremal limit that $\alpha \to \infty$, $r_0\to 0$ with finite charges,
we have
\begin{equation}
M^2_{ext}=Q^2_{D3}+Q^2_{D1}+Q^2_{F},
\end{equation}
which indicates that the bound state is a non-threshold state. The charges
satisfy the relations
\begin{equation}
\frac{Q_{D1}}{Q_{D3}}=\tan\theta, \ \ \ \frac{Q_{F}}{Q_{D3}}
=\frac{\tan\varphi}{\cos\theta}.
\end{equation}
Furthermore, in this bound state, the number of D3-branes is
\begin{equation}
\label{N_3}
N_3=\frac{R_0^4\cos\varphi\cos\theta}{4\pi g_s\alpha'^2},
\end{equation}
the number of D-strings is
\begin{equation}
\label{N_1}
N_1 =\frac{R_0^4\cos\varphi \sin\theta}{4\pi g_s \alpha'^2}
     \frac{V_2}{(2\pi)^2\alpha'},
\end{equation}
and the number of F-strings is
\begin{equation}
\label{N_F}
N_F=\frac{R_0^4\sin\varphi}{4\pi g_s^2 \alpha'^2}\frac{V_2}{(2\pi)^2\alpha'},
\end{equation}
where $R_0^4 =r_0^4\sinh\alpha \cosh\alpha$ and $V_2$ is the area of
the worldvolume coordinates $x_2$ and $x_3$.

We are now going to discuss the various decoupling limits for the supergravity
duals, keeping the number $N_3$ of D3-branes finite.

\subsection{SYM limit}

In this subsection, we first discuss the SYM limit. Taking
the usual decoupling limit:
\begin{equation}
\alpha' \to 0:\ \ \  r=\alpha' u, \ \ \  r_0=\alpha' u_0,
\end{equation}
and keeping $\cos\theta$ and $\cos\varphi$ finite, we have
\begin{equation}
\label{met1}
ds^2= \alpha'\left [\frac{u^2}{\tilde{R}^2}(\cos^{2}\varphi
 (- \tilde{f}dx_0^2 + dx_1^2) +  \cos^{-2}\theta (dx_2^2+dx_3^2))
 + \frac{\tilde{R}^2}{u^2}(\tilde{f}^{-1}du^2 +u^2 d\Omega^2)\right],
\end{equation}
where $\tilde{R}^4=4\pi g_s N_3 \cos\varphi/\cos\theta$ and
$\tilde{f}=1-u_0^4/u^4$. The dilaton, axion and $B$ fields reduce to
\begin{eqnarray}
&& e^{2\phi}= g^2_s \frac{\cos^2\varphi}{\cos^2\theta},\qquad
\chi = 0, 
 \nonumber\\
\label{B}
&& B_{01}= \alpha'^2 \sin\varphi\cos^2 \varphi u^4/{\tilde R}^4, \ \ \
B_{23}=\alpha'^2 \tan\theta u^4 /({\tilde R}^4 \cos^2\theta).
\end{eqnarray}
Obviously, rescaling the closed string coupling constant and worldvolume
coordinates
\begin{equation}
g_s =\frac{\cos\theta }{\cos\varphi}\tilde{g}, \ \
x_{0,1} =\frac{1}{\cos\varphi}\tilde{x}_{0,1}, \ \ x_{2,3}=\cos\theta
  \tilde{x}_{2,3},
\end{equation}
we can convert the metric (\ref{met1}) into a standard form of
$AdS_5\times S^5$:
\begin{equation}
ds^2=\alpha'\left[\frac{u^2}{ {\tilde R}^2}(-\tilde{f}d\tilde{x}_0^2
 +d\tilde{x}_1^2 +d\tilde{x}_2^2 +d\tilde{x}_3^2)
 +\frac{{\tilde R}^2}{u^2}(\tilde{f}^{-1}du^2 +u^2d\Omega_5^2)\right].
\end{equation}
The dilaton, axion and $B$ fields become
\begin{eqnarray}
&& e^{2\phi}=\tilde{g}^2, \qquad
\chi =0, \nonumber\\
&& B_{01}= \alpha'^2 \sin\varphi u^4/{\tilde R}^4, \ \ \
B_{23}=\alpha'^2 \tan\theta u^4 /{\tilde R}^4.
\end{eqnarray}

{}From the above solution, we see that although both the electric and magnetic
components are present in the D3-brane bound state, the resulting theory in
the SYM limit is still the ${\cal N}$=4 SYM with gauge group $U(N_3)$ without
noncommutativity, just as was shown in Ref.~\cite{Lu3}. This can also be
understood from the boundary conditions (\ref{boundary}) of the open string.
In the SYM limit, the mixed boundary conditions reduce to the ordinary
Neumann boundary conditions. The constant $B$ field has no effect on the open
string ending on the D3-branes in that limit. It is worthwhile to stress here
that there is a significant difference between the D3-brane with finite $B$
field and the case without $B$ field depending on whether the D3-brane is
compact or not.

The bound state solution (\ref{solution}) implies that
there is a constant electric component $B_{01}=\sin\varphi$ and a
constant magnetic component $B_{23}=\tan \theta$ of the NS $B$ field on the
worldvolume of the D3-brane, which gives a constant part of the worldvolume
field strength $F_{ij}$. Although the constant part has no effect on the
open string ending on the D3-branes in the SYM limit, it remains in
that limit. If the D3-brane is compact on a torus, this part should be
quantized. We find that the constant part gives $N_1$ units of magnetic flux
and $N_F$ units of electric flux, with $N_1$ and $N_F$ being the numbers of
D-strings and F-strings in the bound state, respectively. Thus the resulting
low energy theory is the ${\cal N}$=4 SYM with both quantized electric and
magnetic fluxes. On the other hand, if the worldvolume of the D3-brane is not
compact, the constant part is physically unmeasurable in the flat infinite
space, as mentioned above~\cite{Gop1}. The resulting theory is then just
the ${\cal N}=4$ SYM as for the case of D3-branes without a $B$ field.

Thus we conclude that in the SYM limit, the worldvolume theory on the
(F1, D1, D3) bound state is the ${\cal N}$=4 SYM without
noncommutativity~\cite{Lu3}.
We will see in the next section that under the $SL(2, {\bf Z})$ transformation
the SYM is mapped again into the SYM with a different coupling constant.

\subsection{NCSYM limit}

Taking the decoupling limit~\cite{Mald2,Has}
\begin{eqnarray}
\alpha' \to 0: && \tan \theta =\frac{\tilde{b}}{\alpha'}, \ \  x_{0,1}
 =\tilde{x}_{0,1}, \ \ x_{2,3}=\frac{\alpha'}{\tilde{b}}\tilde{x}_{2,3},
 \nonumber \\
&& r=\alpha' u,\ \ r_0=\alpha' u_0, \ \ g_s =\alpha' \tilde{g},
\end{eqnarray}
while keeping $\cos\varphi$ finite, we have
\begin{equation}
ds^2 =\alpha' \left[\frac{u^2}{R^2_y} [\cos^2\varphi (-\tilde{f}d\tilde{x}_0^2
 +d\tilde{x}_1^2) +\tilde{h}(d\tilde{x}_2^2 +d\tilde{x}_3^2) ]
  +\frac{R^2_y}{u^2}[\tilde{f}^{-1}du^2 +u^2 d\Omega^2_5]\right],
\end{equation}
where $R^4_y=4\pi \tilde{g}\tilde{b}N_3\cos\varphi$, and
\begin{equation}
\tilde{h}^{-1}=1 +(au)^4, \ \ \  a^4= \tilde{b}^2/R^4_y.
\end{equation}
Also we have
\begin{eqnarray}
&& e^{2\phi} = \tilde{g}^2\tilde{b}^2 \tilde{h}\cos^2\varphi, \qquad
\chi =0, \nonumber\\
&& B_{01}=\alpha'^2 \sin\varphi\cos^2 \varphi u^4/R^4_y, \ \ \
B_{23}=\frac{\alpha'}{\tilde{b}}\frac{(au)^4}{1+(au)^4}.
\end{eqnarray}
After further rescaling the string coupling constant $\tilde{g}$ and the
coordinates $\tilde{x}_{0,1}$ as
\begin{equation}
\tilde{g}\to  \tilde{g}/\cos\varphi,\ \ \ \tilde{x}_{0,1} \to
  \frac{1}{\cos\varphi}\tilde{x}_{0,1},
\end{equation}
we reach
\begin{equation}
\label{met2}
ds^2 =\alpha' \left[\frac{u^2}{R^2} [
 (-\tilde{f}d\tilde{x}_0^2 +d\tilde{x}_1^2) +\tilde{h}(d\tilde{x}_2^2
 +d\tilde{x}_3^2) ]
 +\frac{R^2}{u^2}[f^{-1}du^2 +u^2 d\Omega^2_5]\right],
\end{equation}
and
\begin{eqnarray}
&& e^{2\phi}=\tilde{g}^2\tilde{b}^2 \tilde{h}, \qquad
\chi = 0, \nonumber\\
&& B_{01}=\alpha'^2 \sin\varphi u^4/R^4, \ \  B_{23}=\frac{\alpha'}{\tilde{b}}
 \frac{(au)^4}{1+(au)^4},
\end{eqnarray}
where $a^4=\tilde{b}^2/R^4$ and $R^4= 4\pi \tilde{g}\tilde{b}N_3$.
We note that the geometry (\ref{met2}) is completely the same as in the case
of black D3-branes with only spatial component of $B$ field; for the latter
see Refs.~\cite{Mald2,Has}. It has been claimed that the geometry (\ref{met2})
is the gravity dual configuration of the ${\cal N}$=4 NCSYM with gauge
group $U(N_3)$ and space-space noncommutativity
$[\tilde{x}_2,\tilde{x}_3]=i\tilde{b}$.

In the NCSYM limit, an infinitely
large magnetic field gives rise to the noncommutativity of space-space
while the electric field is kept finite: $F_{01}=\sin\varphi$. The electric
field has no effect on the field theory limit of open string. (It gives rise
to a quantized electric flux if the worldvolume is compact.) We do not find
well-defined field theory limit with space-time noncommutativity. This is
in accordance with the belief that the field theory with space-time
noncommutativity may not be unitary~\cite{GM}. As a result, the low energy
field theory of the bound state (F1, D1, D3) in the NCSYM limit is the
${\cal N}$=4 NCSYM without space-time noncommutativity, and only the spatial
coordinates ($\tilde{x}_{2,3}$) are noncommutative. In addition, we note that
the NCSYM limit implies that $\theta \to \pi/2$ and $\varphi$ is arbitrary
in this case. Hence the decoupling geometry goes to that of the bound state
(F1, D1) with two smeared coordinates as $au \gg 1$.

For low energies, $au \ll 1$, the geometry is that of ordinary SYM,
$AdS_5 \times S_5$. It significantly deviates from this geometry for high
energies, and the deviation appears at the scale of $u\sim 1/a
= R/\sqrt{\tilde b}$.

\subsection{NCOS limit}

The NCOS limit in the dual supergravity is drastically different from
the NCSYM limit, and we study how such a limit can be uniquely determined
in our setting.

In order to get the NCOS limit, we should keep $\Theta^{01}$ finite. This
means that the electric field $e$ should tend to its critical value and
other quantities should scale as given in Eq.~(\ref{ncos}). Note that
$\alpha' G^{ij} \neq 0$ in this limit and the oscillating modes of
an open string do not decouple. This critical behavior is translated in the
supergravity solution~(\ref{solution}) into $\sin\varphi = 1- \alpha'^n e_0/2$
or
\begin{equation}
\cos\varphi=\left(\frac{\alpha'}{\tilde b}\right)^{n/2},
\end{equation}
with $\theta$ kept finite.

Next suppose the scaling behavior of $r$ is given as
\begin{equation}
r=\alpha'^m u.
\end{equation}
We would like to make all our metric in (\ref{solution}) scale as $\alpha'$.
First consider the $dr^2$ term. This is transformed into
\begin{equation}
F^{1/2} dr^2 = \left( 1+ \frac{R^4}{\alpha'^{4m-2}u^4} \right)^{1/2}
\alpha'^{2m} du^2,
\end{equation}
where $R^4 = 4\pi {\tilde g} N_3/({\tilde b}^{n/2}\cos \theta)$ and
$g_s = {\tilde g} \alpha'^{-n/2}$.
We thus see that as long as $m \geq \frac{1}{2}$, this scales as
\begin{eqnarray}
&&\alpha' \left(1+\frac{R^4}{u^4}\right)^{1/2} du^2,
 \;\; {\rm for}\;\; m=\frac{1}{2}, \nonumber \\
&&\alpha' \frac{R^2}{u^2} du^2, \hspace{21mm}{\rm for}\;\; m>\frac{1}{2}.
\label{f}
\end{eqnarray}
Thus all values $m \geq\frac{1}{2}$ are allowed at this point.

We next examine the behavior of other components of the metric.
\begin{eqnarray}
H \sim \frac{{\tilde b}^n}{\alpha'^{4m-2+n}} \frac{R^4}{u^4}, \nonumber\\
G = 1 + \frac{\cos^2 \theta}{\alpha'^{4m-2}} \frac{R^4}{u^4}.
\label{hg}
\end{eqnarray}
{}From Eqs.~(\ref{f}) and (\ref{hg}), we see that all nontrivial functions
of $u$ disappear from the solution for $m>\frac{1}{2}$, and the resulting
metric is $AdS_5\times S_5$ up to a rescaling of the coordinates. This is
believed to correspond to the SYM limit and not the limit we are looking for.
Thus we are uniquely lead to the special scaling $m=\frac{1}{2}$:
\begin{equation}
r= \sqrt{\alpha'} u.
\end{equation}
Note that the parameter $n$ remains arbitrary as long as it is positive.

Having understood how everything scales, we find that the dual gravity
solution corresponding to the NCOS is given by
\begin{equation}
ds^2=\alpha' \tilde{F}^{1/2}\left [ \frac{u^4}{R^4}
 (-\tilde{f} d\tilde{x}_0^2 +d\tilde{x}^2_1) + \frac{1}{\tilde{G}}
 (d\tilde{x}_2^2+d\tilde{x}_3^2 ) +\tilde{f}^{-1}du^2 +u^2d\Omega_5^2\right],
\label{met3}
\end{equation}
where
\begin{equation}
\tilde{F}=1 +\frac{R^4}{u^4}, \ \ \
\tilde{G}=1+\frac{R^4\cos^2\theta}{u^4}.
\label{fg}
\end{equation}
and the coordinates are rescaled as
\begin{eqnarray}
x_{0,1}= \frac{\tilde b^{n/2}}{\alpha'^{(n-1)/2}}\tilde{x}_{0,1},\;\;
x_{2,3}=\sqrt{\alpha'}\tilde{x}_{2,3}.
\end{eqnarray}
The dilaton, axion and $B$ fields are
\begin{eqnarray}
&& e^{2\phi}=\tilde{g}^2 \frac{\tilde{F}^2}{\tilde{G}}\frac{u^4}{\tilde{b}^n
 R^4}, \qquad
\chi = -\frac{{\tilde b}^{n/2}\sin\theta}{{\tilde g}{\tilde F}}, \nonumber\\
&& B_{01}=\alpha' u^4/R^4, \ \  B_{23}=\alpha' \tan\theta/ \tilde{G}.
\end{eqnarray}
We note that the axion field is nonvanishing here, which is quite different
from other solutions.

The geometry (\ref{met3}) is considered to be the supergravity dual of the
NCOS theory with both space-time and space-space noncommutativities. It can
be regarded as an extension of Ref.~\cite{Gop1}, in which the supergravity
dual of NCOS has been given with only space-time noncommutativity by applying
the S-duality to the supergravity dual of NCSYM. In the
NCOS limit, the electric field approaches its critical value while the
magnetic field remains finite. The critical electric field leads to the
space-time noncommutativity. The magnetic field also gives rise to
space-space noncommutativity although it is finite.

Again the geometry is $AdS_5 \times S_5$, that of ordinary SYM for small $u$,
which indicates the low energy limit of NCOS is also the ordinary SYM.
For large $u$, it deviates from this geometry in two ways, one (due to
$\tilde F$) involving space-time coordinates, and the other (due to $\tilde G$)
involving space-space coordinates. These are reflections of the space-time
and space-space noncommutativities, and they arise at the scales of $R$ and
$R\sqrt{\cos\theta}$, respectively.

\sect{$SL(2, {\bf Z})$ duality}

It is well known that IIB superstring has the $SL(2,{\bf Z})$ symmetry
and its low energy approximation, IIB supergravity, has also this symmetry.
It is interesting to consider the relations of the above theories in different
decoupling limits under the $SL(2, {\bf Z})$ transformation.

As it is already pointed out in Ref.~\cite{Gop1}, the S-dual of the NCSYM
gives a NCOS theory without space-space noncommutativity. This case is simple
because the axion field $\chi$ vanishes for the supergravity dual of
NCSYM and the S-duality is achieved just
by taking a simple inverse of the dilaton field. This relation is also
natural because the NCSYM theory is affected only by the magnetic component
of the $B$ field, while the NCOS with $\theta=0$ has only the electric
background. The question we are asking here is what is the general situation.
It might appear that this is again simple, but the above consideration
suggests that these two theories might not be just the S-duals of each other
since the axion field in general does not vanish for the NCOS theory. Indeed
we find that the relation is rather nontrivial.

Here we consider the relations among the three theories discussed
above by using the  more general $SL(2, {\bf Z})$ transformation.
We show that according to the supergravity description, the NCSYM
is always mapped into a NCOS with space-time and space-space
noncommutativities in general.
The converse is rather nontrivial. Under the $SL(2, {\bf Z})$
transformation, we find that the NCOS theory transforms in general into
another NCOS, and for a special case this reduces to a NCSYM theory.

Our IIB supergravity has an $SL(2, {\bf Z})$ invariance as an effective
theory of IIB superstring. The metric in the Einstein frame is inert under
this transformation, so that two solutions after a general $SL(2, {\bf Z})$
transformation are related as
\begin{equation}
ds_E^2 = e^{-\phi/2} ds_{st,1}^2 = e^{-\phi'/2} ds_{st,2}^2,
\end{equation}
where the latter two expressions are two string-frame metrics related
by $SL(2, {\bf Z})$ transformations, and $\phi$ and $\phi'$ are the dilatons
for each solution. Using the notation $\tau = \chi + i e^{-\phi}$,
they are related by
\begin{equation}
\tau' =\frac{a \tau + b}{c \tau + d}, \;\; ad-bc =1,
\end{equation}
for integers $a,b,c$ and $d$. (The usual S-duality corresponds to the case
$\tau' =-1/\tau$.) This gives us
\begin{equation}
e^{-\phi'} = \frac{e^{-\phi}}{|c\tau +d|^2},
\end{equation}
resulting in
\begin{equation}
ds_{st,2}^2 = |c \tau+ d| ds_{st,1}^2.
\label{sdual}
\end{equation}

Let us first discuss the $SL(2,{\bf Z})$ transformation of the SYM
theory~(\ref{met1}). Since the axion field vanishes and the dilaton is
a constant, the supergravity dual $AdS_5\times S^5$ is still of the form
$AdS_5\times S^5$ after the $SL(2,{\bf Z})$ transformation. This implies
that the ordinary SYM theory is mapped into itself again, but with a
different coupling constant
\begin{equation}
{\tilde g}' = {\tilde g}(d^2 + c^2 {\tilde g}^{-2}).
\end{equation}

Next, for the NCSYM~(\ref{met2}), the axion is again zero and
\begin{equation}
\tau=i/{\tilde h}^{1/2}{\tilde g}{\tilde b}.
\end{equation}
The $SL(2, {\bf Z})$ transformation then leads to
\begin{eqnarray}
\label{met21}
ds^2 &=&\alpha' \tilde{h}'{}^{-1/2}
 \left[\frac{u^2}{R^2} [ (-\tilde{f}d\tilde{x}_0^2 +d\tilde{x}_1^2)
 +\tilde{h}(d\tilde{x}_2^2 +d\tilde{x}_3^2) ]
 +\frac{R^2}{u^2}[f^{-1}du^2 +u^2 d\Omega^2_5]\right], \nonumber\\
{\tilde h}'{}^{-1} &=& d^2 + \frac{c^2}{{\tilde g}^2{\tilde b}^2}
 {\tilde h}^{-1}, \qquad
e^{2\phi'} = {\tilde g}^2{\tilde b}^2{\tilde h}{\tilde h}'{}^{-2}.
\end{eqnarray}
This metric can be cast into the form (\ref{met3}) with
\begin{equation}
\cos^2 \theta' = \frac{c^2}{c^2 + (d {\tilde b}{\tilde g})^2},
\end{equation}
up to a coordinate and parameter rescaling. This means that the NCSYM
transforms into a NCOS with both space-time and space-space noncommutativities
after the transformation. When $d=0$ (the case of S-duality), the
space-space noncommutativity then disappears.

Finally let us consider the NCOS theory~(\ref{met3}) with both space-time
and space-space noncommutativities where the axion field does not vanish.
In the above two cases, the simple S-duality $\tau \to -\frac{1}{\tau}$ was
useful to get information on their relations, but here we consider the
$SL(2, {\bf Z})$ transformation with
\begin{equation}
\tau = -\frac{{\tilde b}^{n/2}\sin\theta}{{\tilde g}{\tilde F}}
 + i \frac{{\tilde b}^{n/2}{\tilde G}^{1/2}R^2}{{\tilde g}{\tilde F}u^2}.
\end{equation}
The factor in Eq.~(\ref{sdual}) then becomes
\begin{eqnarray}
|c\tau + d|= \left[ ( c{\tilde b}^{n/2} \sin\theta - d{\tilde g}{\tilde F})^2
 + c^2 {\tilde b}^n {\tilde G}\frac{R^4}{u^4}\right]^{1/2}
 \frac{1}{{\tilde g}{\tilde F}},
\end{eqnarray}
which, with the help of Eq.~(\ref{fg}), is transformed into
\begin{equation}
\frac{{\hat F}^{1/2}}{{\tilde F}^{1/2}},
\label{ncos1}
\end{equation}
where
\begin{equation}
{\hat F} \equiv \left(d - \frac{c{\tilde b}^{n/2}}{\tilde g} \sin\theta
\right)^2 + \left(d^2 + \frac{c^2 {\tilde b}^{n}}{{\tilde g}^2} \cos^2 \theta
\right)\frac{R^4}{u^4}.
\label{sncos}
\end{equation}
When Eq.~(\ref{ncos1}) is used in Eqs.~(\ref{sdual}) and (\ref{met3}), we find
that the $SL(2, {\bf Z})$-transformed solution is the same as the original
one (\ref{met3}) with $\tilde F$ replaced by $\hat F$. The new $\theta'$ is
given by
\begin{equation}
\cos^2 \theta' = \frac{(d{\tilde g} - c{\tilde b}^{n/2}\sin\theta)^2}
 {d^2{\tilde g}^2 + c^2 {\tilde b}^n \cos^2 \theta}\cos^2 \theta.
\end{equation}

We thus find that a NCOS theory transforms into another NCOS theory with a
different noncommutativity parameter under the $SL(2,{\bf Z})$ transformation.
In general both the NCOSs have space-time and space-space noncommutativities.
However, in a special case when the first term in Eq.~(\ref{sncos}) vanishes,
\begin{equation}
d - \frac{c{\tilde b}^{n/2}}{\tilde g} \sin\theta =0,
\end{equation}
we find  that the $SL(2, {\bf Z})$-transformed solution reduces to the
form ({\ref{met2}), which describes a NCSYM. This is possible only for the
case in which the asymptotic value of the axion
$\chi_\infty = -\frac{{\tilde b}^{n/2}\sin\theta}{\tilde g}$ is a rational
number, in agreement with the conclusion in Ref.~\cite{RS} derived from the
open string point of view. This result is also consistent with that in Ref.
\cite{LRS}.\footnote{If we used the classical $SL(2,{\bf R})$
symmetry in the original solution~(\ref{solution}), we could shift the
asymptotic value $\chi_\infty$ of the axion. In particular, we could set
it to zero. This is the case discussed in Ref.~\cite{LRS}; the NCOS transforms
into a NCSYM after the S-duality. It was argued there that the S-dual of
NCSYM is not always NCOS, but that limit is different from the
well-defined NCOS limit.} This also includes the case $\theta=0$
discussed in Ref.~\cite{Gop1}. In this case one can also reach the same
conclusion by using a simple S-duality.

\sect{Conclusions and discussions}

In this paper, we have considered in the framework of IIB supergravity three
different scaling limits for the bound state (F1, D1, D3) and
$SL(2,{\bf Z})$ transformations of the resulting theories. These three
theories are, respectively, (1) the ordinary ${\cal N}$=4 SYM with or without
$N_F$ units of electric flux and $N_1$ units of magnetic flux depending on
whether the worldvolume of the D3-brane is compact or not (here $N_1$ and $N_F$
are, respectively, the numbers of D-strings and F-strings in the bound state);
(2) the NCSYM with or without $N_F$ units of electric flux; and (3) the NCOS
with both space-time and space-space noncommutativities. Under a general
$SL(2,{\bf Z})$ transformation, the gravity dual $AdS_5\times S^5$ of
the SYM still has the same form up to the rescaling of parameters and
coordinates. This implies that the SYM becomes another SYM with different
coupling constants after the transformation. The gravity dual of NCSYM
takes the form of the NCOS in general with both the space-time and
space-space noncommutativities. This means that a NCSYM transforms into
a NCOS under the $SL(2,{\bf Z})$ transformation. Finally, the
gravity dual of NCOS with both space-time and space-space noncommutativities
remains in the same form after the $SL(2,{\bf Z})$ transformation. Our
result implies that in general a NCOS transforms into another NCOS with
different noncommutativity parameter. In the special case when
the asymptotic value of the axion is a rational number, it is possible to
transform a NCOS into a NCSYM, and when the asymptotic value vanishes, one
can reach this conclusion by using a simple S-duality alone.

As mentioned before, the thermodynamic quantities (\ref{thermo}) are
independent of the parameter angles $\varphi$ and $\theta$. This implies
that the thermodynamics is the same for three different theories, SYM, NCSYM,
and NCOS, in this supergravity approximation. Also the thermodynamics
remains unchanged under the $SL(2,{\bf Z})$ transformation, since the latter
does not change the form of Einstein metric.

{}From the solution (\ref{solution}) we see that when $\theta =\pi/2$ and
$\varphi$ is arbitrary, the bound state solution goes to that for
the (F1, D1) bound state. Taking the critical electric field limit, one
has a 2-dimensional NCOS from the bound state (F1, D1)~\cite{Gop1,Kleb}.
{}From our previous discussions on the $SL(2,{\bf Z})$ transformation, one
can see that in general the 2-dimensional NCOS becomes another 2-dimensional
NCOS with different noncommutativity parameters, as in the case of
4 dimensions. When the asymptotic value of the axion is a rational number,
it is possible to transform the 2-dimensional NCOS into an ordinary
2-dimensional SYM with  quantized electric flux~\cite{Kleb}. Of course,
when the asymptotic value of the axion vanishes, one can transform
the 2-dimensional NCOS into a 2-dimensional SYM only by using S-duality,
which is precisely the case in Ref.~\cite{Kleb}.

An interesting extension of our work is to consider more general solutions
like (F1, D1, NS5, D5) bound states and their scaling limits. We suspect
that in the bound state, SYM, NCSYM, NCOS and the so-called little string
theory are connected with each other through the $SL(2,{\bf Z})$
transformation. This conjecture is currently under investigation.

\section*{Acknowledgements}

We would like to thank A. Fujii, J.X. Lu and N. Yokoi for valuable discussions.
This work was supported in part by the Japan Society for the Promotion
of Science and in part by Grants-in-Aid for Scientific Research Nos. 99020
and 12640270, and by a Grant-in-Aid on the Priority Area: Supersymmetry and
Unified Theory of Elementary Particles.

\end{document}